\documentstyle[epsf]{mn}

\newif\ifAMStwofonts

\def\arcdeg{^{\circ}}
\def\apj{ApJ}
\def\apjl{ApJL}
\def\apjs{ApJS}
\def\mnras{MNRAS}
\def\aap{A\&A}
\def\aaps{A\&AS}
\def\procspie{Proc.~SPIE}

\def\pasj{PASJ}

\ifoldfss
  \ifCUPmtlplainloaded \else
    \NewTextAlphabet{textbfit} {cmbxti10} {}
    \NewTextAlphabet{textbfss} {cmssbx10} {}
    \NewMathAlphabet{mathbfit} {cmbxti10} {} 
    \NewMathAlphabet{mathbfss} {cmssbx10} {} 
  \fi
  \ifAMStwofonts
    \ifCUPmtlplainloaded \else
      \NewSymbolFont{upmath} {eurm10}
      \NewSymbolFont{AMSa} {msam10}
      \NewMathSymbol{\upi}     {0}{upmath}{19}
      \NewMathSymbol{\umu}     {0}{upmath}{16}
      \NewMathSymbol{\upartial}{0}{upmath}{40}
      \NewMathSymbol{\leqslant}{3}{AMSa}{36}
      \NewMathSymbol{\geqslant}{3}{AMSa}{3E}

    \fi
  \fi
\fi 

\ifnfssone
  \newmathalphabet{\mathit}
  \addtoversion{normal}{\mathit}{cmr}{m}{it}
  \addtoversion{bold}{\mathit}{cmr}{bx}{it}
  \newmathalphabet{\mathbfit} 
  \addtoversion{normal}{\mathbfit}{cmr}{bx}{it}
  \addtoversion{bold}{\mathbfit}{cmr}{bx}{it}
  \newmathalphabet{\mathbfss} 
  \addtoversion{normal}{\mathbfss}{cmss}{bx}{n}
  \addtoversion{bold}{\mathbfss}{cmss}{bx}{n}
  \ifAMStwofonts
    \ifCUPmtlplainloaded \else
      \UseAMStwoboldmath
      \makeatletter
      \new@mathgroup\upmath@group
      \define@mathgroup\mv@normal\upmath@group{eur}{m}{n}
      \define@mathgroup\mv@bold\upmath@group{eur}{b}{n}
      \edef\UPM{\hexnumber\upmath@group}
      \new@mathgroup\amsa@group
      \define@mathgroup\mv@normal\amsa@group{msa}{m}{n}
      \define@mathgroup\mv@bold\amsa@group{msa}{m}{n}
      \edef\AMSa{\hexnumber\amsa@group}
      \makeatother
      \mathchardef\upi="0\UPM19
      \mathchardef\umu="0\UPM16
      \mathchardef\upartial="0\UPM40
      \mathchardef\leqslant="3\AMSa36
      \mathchardef\geqslant="3\AMSa3E
    \fi
  \fi
\fi 

\ifnfsstwo
  \DeclareMathAlphabet{\mathbfit}{OT1}{cmr}{bx}{it}
  \SetMathAlphabet\mathbfit{bold}{OT1}{cmr}{bx}{it}
  \DeclareMathAlphabet{\mathbfss}{OT1}{cmss}{bx}{n}
  \SetMathAlphabet\mathbfss{bold}{OT1}{cmss}{bx}{n}
  \ifAMStwofonts
    \ifCUPmtlplainloaded \else
      \DeclareSymbolFont{UPM}{U}{eur}{m}{n}
      \SetSymbolFont{UPM}{bold}{U}{eur}{b}{n}
      \DeclareSymbolFont{AMSa}{U}{msa}{m}{n}
      \DeclareMathSymbol{\upi}{0}{UPM}{"19}
      \DeclareMathSymbol{\umu}{0}{UPM}{"16}
      \DeclareMathSymbol{\upartial}{0}{UPM}{"40}
      \DeclareMathSymbol{\leqslant}{3}{AMSa}{"36}
      \DeclareMathSymbol{\geqslant}{3}{AMSa}{"3E}
    \fi
  \fi
\fi 

\ifCUPmtlplainloaded \else
  \ifAMStwofonts \else 
    \def\upi{\pi}
    \def\umu{\mu}
    \def\upartial{\partial}
  \fi
\fi

\title[Accretion column eclipses]{Accretion column eclipses in the X-ray
  pulsars GX~1+4 and RX~J0812.4$-$3114}
\author[D. K. Galloway et al.]
  {D.~K. Galloway,$^{1,2,3}$ A.~B. Giles,$^{1,4}$ K. Wu,$^{2,5}$ and
    J.~G. Greenhill$^{1}$ \\
  $^1$ School of Mathematics and Physics, University of Tasmania,
       Hobart 7001, Australia \\
  $^2$ RCfTA, School of Physics, University of Sydney, NSW 2006, 
Australia\\
  $^3$ present address: Center for Space Research, MIT, 37-571 
       77 Massachusetts Avenue, Cambridge, MA 02139 \\
  $^4$ visitor, USRA/Laboratory for High Energy Astrophysics, Goddard 
Space
       Flight Center, Maryland \\
  $^5$ MSSL, University College London, Holmbury St. Mary, Dorking, Surrey
       RH5 6NT, UK}

\pagerange{\pageref{firstpage}--\pageref{lastpage}}
\pubyear{2000}

\begin{document}

\maketitle

\label{firstpage}

\begin{abstract}
Sharp dips observed in the pulse profiles of three X-ray pulsars (GX~1+4,
RX~J0812.4$-$3114 and A~0535+26) have previously been suggested to arise
from partial eclipses of the emission region by the accretion column
occurring once each rotation period.  We present pulse-phase spectroscopy
from {\it Rossi X-ray Timing Explorer\/} satellite observations of GX~1+4
and RX~J0812.4$-$3114 which for the first time confirms this
interpretation. The dip phase corresponds to the closest approach of the
column axis to the line of sight, and the additional optical depth for
photons escaping from the column in this direction gives rise to both the
decrease in flux and increase in the fitted optical depth measured at this
phase.  Analysis of the arrival time of individual dips in GX~1+4 provides
the first measurement of azimuthal wandering of a neutron star accretion
column. The column longitude varies stochastically with standard deviation
2--6$\arcdeg$ depending on the source luminosity.  Measurements of the
phase width of the dip both from mean pulse profiles and individual
eclipses demonstrates that the dip width is proportional to the flux.  The
variation is consistent with that expected if the azimuthal extent of the
accretion column depends only upon the Keplerian velocity at the inner
disc radius, which varies as a consequence of the accretion rate
$\dot{M}$.
\end{abstract}


\section{Introduction}

\begin{figure}
 \epsfxsize=9.0cm 
 \epsfbox{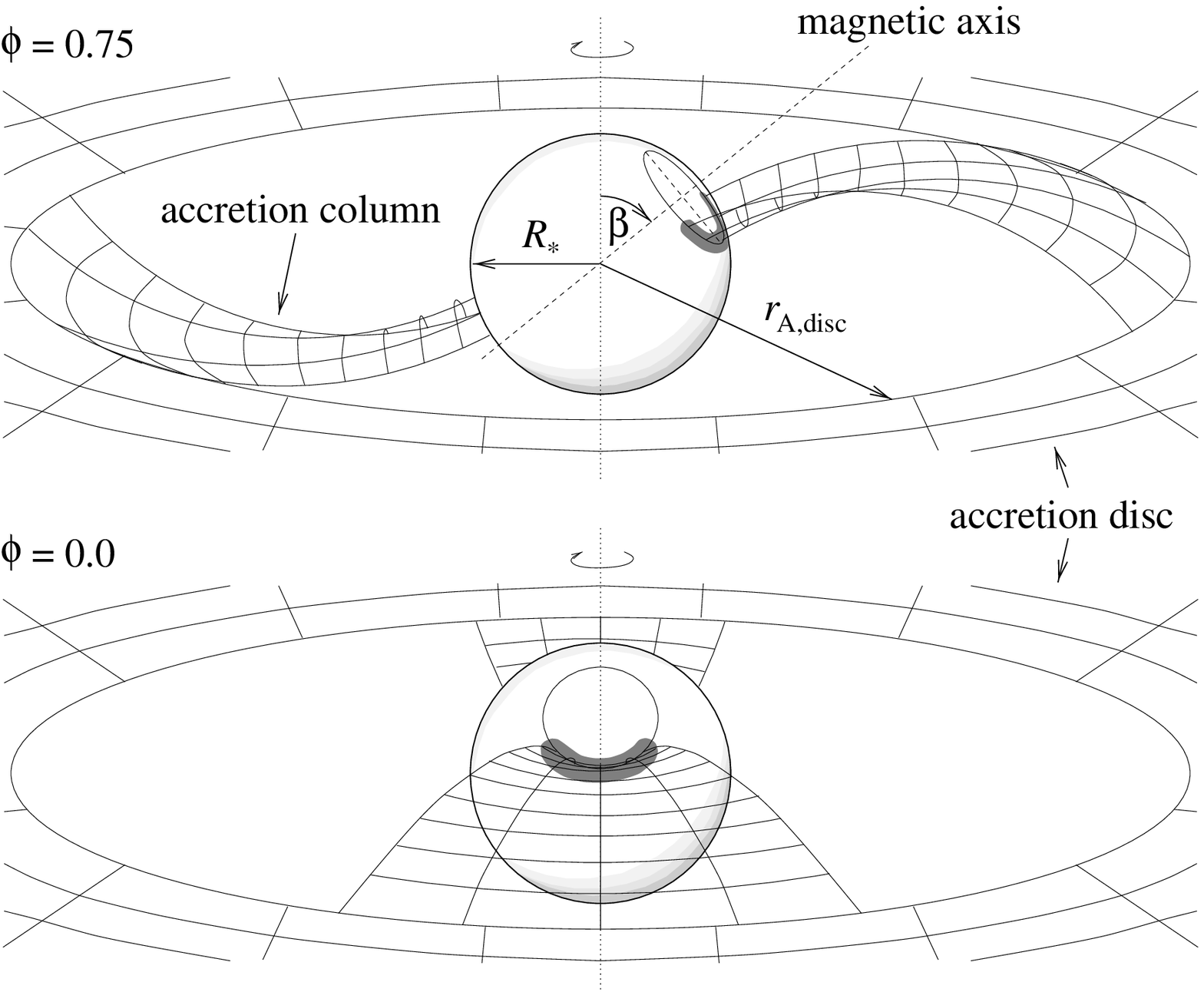}
 \caption[nature1.eps]{ Schematic showing the accretion geometry of X-ray
pulsars (not to scale). The accreting gas from the companion is thought to
form an accretion disc far from the neutron star, which is disrupted at
$r_{\rm A,disc}$ due to the strong (dipolar) magnetic field. Within this
radius the accreting matter may only flow along the magnetic field lines,
forming two accretion columns terminating at the magnetic poles. Thermal
X-ray emission originates from the shaded region at each pole and is
Compton scattered by the accretion column plasma above it.  Here the
magnetic axis is aligned relative to the neutron star rotation axis by
$\beta$ (the magnetic colatitude). The inclination angle $i$ is such that
at phase $\phi=0.0$ one of the accretion columns will be closely aligned
with the line of sight, giving rise to a partial `eclipse' of the shaded
emission region by the accretion column.
 \label{schem} }
\end{figure}

Binary X-ray pulsars are neutron stars which accrete stellar material from
a companion star \nocite{white95}(e.g. {White}, {Nagase} \& {Parmar} 1995). The accretion flow is typically
mediated through an accretion disc and accretion columns delineated by the
magnetic field of the neutron star ($B\sim10^{12}$~G at the surface) which
terminate at strongly-emitting regions near the magnetic poles
\nocite{gl79a,gl79b}(e.g. {Ghosh} \& {Lamb} 1979a,b).  The plasma flows along the columns from the
inner edge of the accretion disc, where the magnetic field begins to
dominate the accretion flow, to the magnetic poles (Fig. \ref{schem}).
The base of the columns are expected to cover an arc-shaped region on the
neutron star surface as a consequence of the dipolar magnetic field and
the limited set of magnetic field lines along which accretion is
energetically favourable.  To date, observational verification of this
scenario has not been possible due to instrumental limitations and a lack
of theoretical models describing the anisotropy of X-ray emission from the
column.  Optical spectroscopic and polarimetric measurements of accreting
white dwarf stars in cataclysmic variables \nocite{warner95}(e.g. {Warner} 1995) provide
corroboration, but such techniques are not available for X-ray pulsars
where the accretion columns radiate in X-rays rather than the optical and
UV bands.  In general there is little observational data available which
allow detailed studies of the dynamics of accretion flows.

Sharp (phase width $\Delta\phi\approx0.1$) dips are usually observed in
the pulse profiles of the X-ray pulsars A~0535+26, RX~J0812.4$-$3114 and
GX~1+4.  These dips have previously been attributed to interactions of the
accretion column and the emission region \nocite{cem98,rei99,gil00}({\v{C}emelji\'c} \& {Bulik} 1998; {Reig} \& {Roche} 1999; {Giles} {et~al.} 2000)
although supporting arguments have been scarce.  A~0535+26 is observed
only in outburst, when the X-ray emission is modulated on a period of
$P_{\rm spin}\approx103.5$~s \nocite{neg00}(e.g. {Negueruela} {et~al.} 2000).
RX J0812.4$-$3114, the X-ray counterpart to the B0-1 III-IV star LS~992
was identified from cross-correlation of {\it ROSAT} galactic plane survey
data with SIMBAD OB star catalogues \nocite{motch97}({Motch} {et~al.} 1997).  Hourly flux
variations, as well as X-ray pulsations at $P\approx31.9$~s were
discovered during a {\it Rossi X-ray Timing Explorer} ({\it RXTE\/})
observation of the source on 1998 February
\nocite{rei99}({Reig} \& {Roche} 1999).  The X-ray behaviour of these Be/X-ray binaries is
strongly dependent upon the orbital parameters, and the neutron star spin
periods $P_{\rm spin}$ show a strong correlation with the orbital period
$P_{\rm orb}$ \nocite{corbet86}({Corbet} 1986).  The $\approx81$~d periodicity suggested
by {\it RXTE\/} All-Sky Monitor (ASM) data obtained between 1998--1999
\nocite{corbet00}({Corbet} \& {Peele} 2000) is consistent with this correlation and probably
represents the binary period.  The distance was estimated at 9~kpc based
on the companion spectral type and interstellar reddening, with a likely
uncertainty of 50 per cent \nocite{motch97}({Motch} {et~al.} 1997).  The pulse profile from the
source is strongly asymmetric, with a sharp dip forming the primary
minimum.  The energy spectrum measured by {\it RXTE}\/ was fairly typical
for X-ray pulsars, and could be adequately fit using a power law model
with exponential cutoff \nocite{rei99}({Reig} \& {Roche} 1999).

GX~1+4 is the most persistently bright of the three, and posesses the
slowest period.  Measurements by {\it RXTE\/} and the Burst and Transient
Source Experiment (BATSE) aboard the {\it Compton Gamma-Ray
Observatory} \nocite{zhang95}({Zhang} {et~al.} 1995) between 1996--7 found $P_{\rm
spin}=123.5$--125.5~s (Galloway \& Greenhill 2001, in preparation).  The
companion to the X-ray source is the 17th magnitude M6 giant V2116
Ophiuchus \nocite{dav77}({Davidsen}, {Malina} \& {Bowyer} 1977), the only known symbiotic binary with a neutron
star as the energy source for the emission line nebula \nocite{symcat00}({Belczy{\'n}ski} {et~al.} 2000).
Regular episodes of relatively rapid spin-up measured from BATSE
monitoring of the source suggest an orbital period of $\approx 304$~d
\nocite{per99}({Pereira}, {Braga} \& {Jablonski} 1999). The distance to the source is thought to be 3--15~kpc
\nocite{chak97:opt}({Chakrabarty} \& {Roche} 1997).  Several estimates of the magnetic field strength are
in the range 2--3$\times 10^{13}$~G
\nocite{beu84,dot89,mony91,gre93,cui97}({Beurle} {et~al.} 1984; {Dotani} {et~al.} 1989; {Mony} {et~al.} 1991; {Greenhill} {et~al.} 1993; {Cui} 1997).  The dips in this extremely
variable source are clearly visible in the X-ray lightcurve when the
source is bright and the time resolution is sufficient, and are present in
the folded pulse profiles even when the luminosity falls almost to the
detection limit \nocite{gil00}({Giles} {et~al.} 2000). When the source is bright the dips are
broad and generally do not reach zero, while at lower luminosities they
are narrower and emission at the dip phase is consistent with the
background level.  The phase-averaged X-ray spectra from the source are
consistent with an approximately thermal spectrum (with temperature
$T_0\approx1.3$~keV) modified by Compton scattering within hotter material
($T_{\rm e}=6$--10~keV) with characteristic optical depth $\tau=2$--6
\nocite{myphd:a}({Galloway} 2000b).  The thermal spectrum is thought to arise from the
region of the polar cap at the base of the accretion column, with Compton
scattering occurring in the hotter plasma above.

 \section{Observations and analysis}

We undertook a detailed study of the phenomenology of dips in GX~1+4, and
to a lesser extent RX~J0812.4$-$3114, using archival {\it RXTE\/}
satellite observations.  Data were obtained using the Proportional Counter
Array \nocite{xte96}(PCA; {Jahoda} {et~al.} 1996), which is sensitive to X-ray photons in the
energy band 2--60~keV.  Six of the eight event analysers (EAs) aboard {\it
RXTE\/} are available for processing of events measured by the PCA. Two
EAs are dedicated to modes which are always present, `Standard-1' and
`Standard-2'.  The Standard-1 mode features 0.25-$\mu{\rm s}$ time
resolution but only one spectral channel, while Standard-2 offers 128
spectral channels between 0--100~keV (the sensitivity above 60~keV is
negligible) accumulated every 16~s.  GoodXenon mode data, which offers the
maximum 256 channel spectral resolution on 0.9537-$\mu{\rm s}$ time
resolution, was available for most of the observations.  Analysis of {\it
RXTE\/} data presented here was carried out using {\sc lheasoft~5.0},
released 23 February 2000 by the {\it RXTE\/} Guest Observer Facility
(GOF). Spectral response matrices were calculated for each observation in
order to correct for gain changes over the observation period. A separate
matrix was calculated for each of the three PCA layers using {\sc pcarmf},
and then summed using {\sc addrmf}; this was recommended by the PCA team
as a workaround to a minor bug in {\sc pcarmf} which occurs when
calculating the response for all layers simultaneously (see
http://heasarc.gsfc.nasa.gov/docs/xte/xhp\_new.html\#bug)

Observations of GX~1+4 were made in 35 separate intervals
between 1996 February and 1997 May, with a total exposure time of 230~ks.
For a detailed description of the observations and data processing see
\nocite{myphd:a}{Galloway} (2000b).  The pulse period for each observation was determined
through folding 1~s lightcurves (with times corrected to the solar-system
barycentre) on a range of periods $P_{\rm trial}$ close to the expected
value and calculating the resulting $\chi^2$. The most probable period was
found from fitting the peak in the $\chi^2$ versus $P_{\rm trial}$ plot
using a gaussian, with errors determined from equivalent analysis on 100
simulated lightcurves.  During 1996--1997 GX~1+4 was found to be spinning
down at a relatively steady rate, and the pulse period increased from
123.6~s to 125.6~s in the year following 1996 February. The periods
measured by {\it RXTE\/} are consistent with those from BATSE
\nocite{chak97}({Chakrabarty} {et~al.} 1997) except when the source was very faint (e.g. 1996 September
through November). At these times the the BATSE estimates appear
unreliable, exhibiting (apparently) significant variations of $\pm0.5$~s
on timescales of $\approx10$~d. The periods measured by $P_{\rm trial}$
folding of {\it RXTE\/} data on the other hand increase monotonically
throughout this interval.

Two observations of RX~J0812.4$-$3114 were made by {\it RXTE\/} on 1998
February 1 and 3 and one on 1999 March 25.  The 1999 observation was
scheduled at the expected time for an outburst, which have occurred
regularly on an $\approx81$~d period since 1998 \nocite{corbet00}({Corbet} \& {Peele} 2000).  The
measured period for RX~J0812.4$-$3114 was $P=31.885623(9) \pm
0.000007(5)$~s, in excellent agreement with the value
$P=31.8856\pm0.0001$~s obtained by \nocite{corbet00}{Corbet} \& {Peele} (2000) for the same
observation.  Previous spectral analyses for this source used a power-law
model with an exponential cutoff \nocite{rei99,corbet00}({Reig} \& {Roche} 1999; {Corbet} \& {Peele} 2000).  Using the
recently updated response matrices and faint background models, the cutoff
power law model results in generally unacceptable fits, as do power law,
power law and blackbody, and broken (two index) power law. Instead the
best--fitting spectral model, as for the GX~1+4 observations, was the
Comptonisation continuum component of \nocite{tit94}{Titarchuk} (1994). This analytic
approximation is implemented in {\sc xspec} as `{\tt compTT}' and resulted
in a maximum $\chi^2_{\nu}=1.5$ for the 1999 March 25 observation.
Spectral fit parameters over the three observations were $T_0=1$--1.2~keV,
$T_{\rm e}=5.2$--6.4~keV and $\tau=3.8$--4.6.  In contrast to GX~1+4, no
gaussian model component (representing Fe line emission at 6.4--6.7~keV)
is required for the spectral fits.  The neutral column density $n_{\rm H}$
was typically consistent with zero to within the $1\sigma$ uncertainties.
The luminosity in the 2--60~keV energy range reached a maximum during the
1999 March 25 observation of $(2.57\pm0.33)\times10^{36}\ {\rm
erg\,s}^{-1}$ (for a source distance of 9~kpc).

\subsection{Pulse-phase spectroscopy}

The GoodXenon mode data from the 1999 March 25 {\it RXTE\/} observation of
RX~J0812.4$-$3114 and selected observations of GX~1+4 were divided into 10
equal phase bins, and a spectrum obtained for each phase range.  For
GX~1+4 the observations on 1996 February 17 and June 8 were chosen for
their similar fitted values of $n_{\rm H}$ to minimise any potential
effects due to variation in this parameter.  Additionally, they represent
examples of the two apparently distinct groups of GX~1+4 spectra.  When
$L_{\rm X}\ga2\times10^{37}\ {\rm erg\,s}^{-1}$ (2--60~keV, assuming a
source distance of 10~kpc), $\tau=4$--6 and $T_{\rm e}=6$--9~keV
typically, while at lower $L_{\rm X}$ $\tau\approx3$ and $T_{\rm
e}=7$--14~keV \nocite{myphd:a}({Galloway} 2000b). The 1996 February 17 observation
corresponds to the former case, with $L_{\rm X}=(8.5\pm3.3)\times10^{37}\
{\rm erg\,s}^{-1}$, $\tau=5.6\pm0.1$ and $T_{\rm e}=8.1\pm0.1$~keV,
while on 1996 June 8 $L_{\rm X}=(1.33\pm0.54)\times10^{37}\ {\rm
erg\,s}^{-1}$, $\tau=2.5_{-1.1}^{+0.5}$ and $T_{\rm
e}=13.6_{-2.6}^{+5.1}$~keV ($1\sigma$ confidence limits). The measured
pulse periods during the two observations were $123.6368(1) \pm 0.0003(8)$
and $124.1677(0) \pm 0.0005(6)$~s respectively.  The ephemeris was chosen
so that the primary minimum falls at phase 0.0, with the first bin centred
there.  Each of the 10 spectra were then fitted with a model identical to
that used for the phase averaged spectra, with a Comptonisation component
(`{\tt compTT}') and a gaussian component (in the case of GX~1+4) both
attenuated by neutral interstellar material.

\begin{figure}
 \epsfxsize=9.0cm 
 \epsfbox{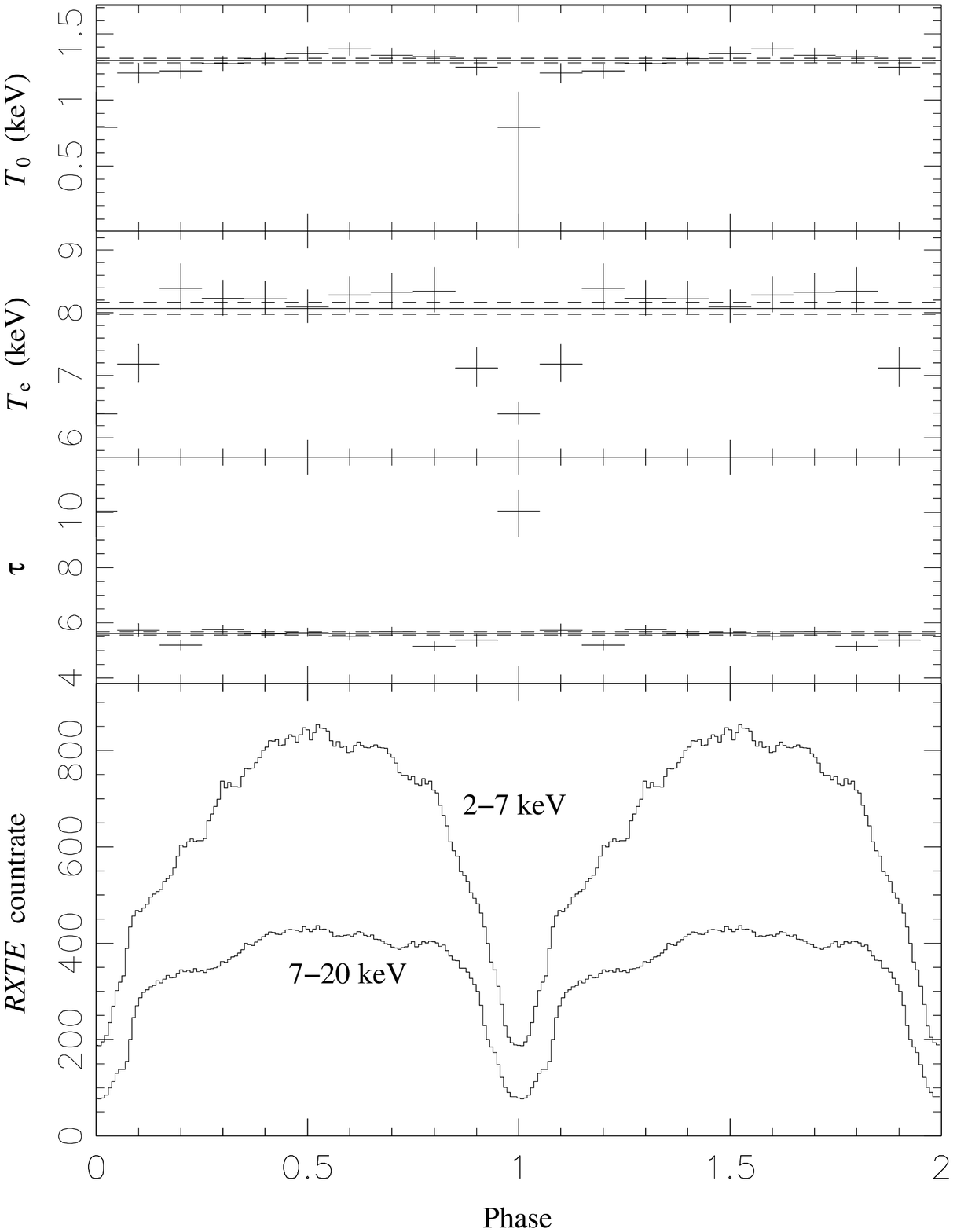}
 \caption[B-pps.eps]{ Comptonisation component fit parameters versus
pulse phase for the 1996 February 17 {\it RXTE\/} observation of GX~1+4.
The pulse period used is $P=123.63681$~s; the ephemeris is
defined arbitrarily such that the dip in the pulse profile falls at phase
0.0. From top to bottom, the panels shows the source spectrum temperature
$T_0$, scattering plasma temperature $T_{\rm e}$ and optical depth $\tau$
and the integrated pulse profile in 2--7 and 7--20~keV energy bands
respectively. Fitted parameter values for the phase-averaged spectra
covering the same interval are shown by the solid lines; error bars and the
dotted lines show the 1$\sigma$ confidence intervals.  Two full pulse periods
are shown for clarity. The total observing time was 10400~s.
\label{B-pps} }
\end{figure}

\begin{figure}
 \epsfxsize=9.0cm 
 \epsfbox{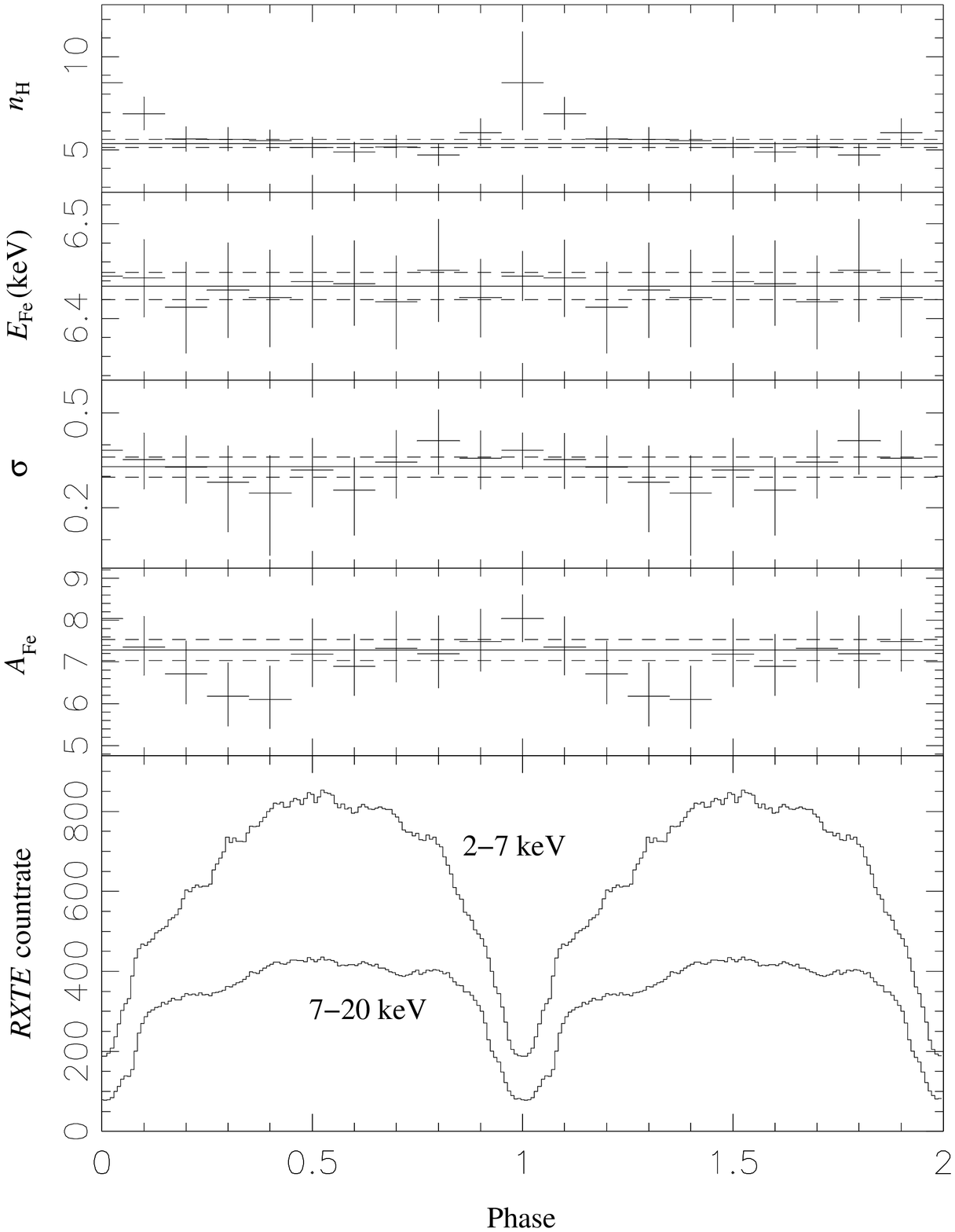}
 \caption[B-pps-2.eps]{ Additional spectral model fit parameters versus
pulse phase for the 1996 February 17 {\it RXTE\/} observation of GX~1+4.
From top to bottom, the panels shows the neutral column density $n_{\rm
H}$ (in units of $10^{22}\ {\rm cm}^{-2}$), Fe line component centre
energy $E_{\rm Fe}$, width $\sigma_{\rm Fe}$ and normalisation $A_{\rm
Fe}$ (units of $10^{-3}$~photons$\,{\rm cm^{-2}\,s^{-1} \,keV^{-1}}$) and
the pulse profiles in the 2--7 and 7--20~keV energy bands. Other details
are as for Fig.  \ref{B-pps}.
 \label{B-pps-2} }
\end{figure}

The greatest spectral variation with phase is typically observed close to
the primary minimum (Fig. \ref{B-pps}). For the 1996 February 17
observation of GX~1+4, the source spectrum temperature $T_0$ is generally
quite consistent with the phase-averaged value, except at phase $\phi=0.0$
where it drops to $\approx0.75$~keV. The scattering plasma temperature is
also lower at $\phi=0.0$ as well as the two adjacent bins, by at most
$\approx20$ per cent (a 7.5$\sigma$ deviation from the phase- averaged
value). The optical depth $\tau$ exhibits a dramatic increase at the phase
of primary minimum, from around 5.6 for the phase-averaged spectrum to
more than 10 (4.7$\sigma$).   The {\tt compTT} normalisation (not shown)
exhibits the greatest variation of all the parameters, and traces the
pulse profile quite well. The 1$\sigma$ confidence limits on the
parameters were generally small except at $\phi=0.0$. The spectral
variation was symmetric about the primary minimum, as was the pulse
profile.  The column density $n_{\rm H}$ was $\approx 60$\% greater than
the phase-averaged spectral value around $\phi=0.0$ (Fig. \ref{B-pps-2}),
but at a significance level of less than 2$\sigma$. Variation in the Fe
line component parameters was no greater than 1.5$\sigma$ with respect to
the phase-averaged values over the entire phase range.

The phase variation of spectral fit parameters for the 1996 June 8
observation (not shown) was less significant due to the lower countrate
and the wider confidence limits that result. Since the gaussian component
parameters were shown to exhibit no significant variation with phase when
the source was much brighter, for this analysis they were fixed to values
found for the phase-averaged spectrum.  Reliable limits on the plasma
temperature $T_{\rm e}$ could only be determined for phases $\phi=0.0$ and
0.1; in the other phase bins the value was instead frozen at the value
obtained for the phase-averaged spectrum.  The optical depth was greater
compared to the phase-averaged level in the phase bins centred on
$\phi=0.0$ and 0.1.  While the maximum enhancement is $\approx 70$\% the
significance is only $1.5\sigma$ at best.

\begin{figure}
 \epsfxsize=9.0cm 
 \epsfbox{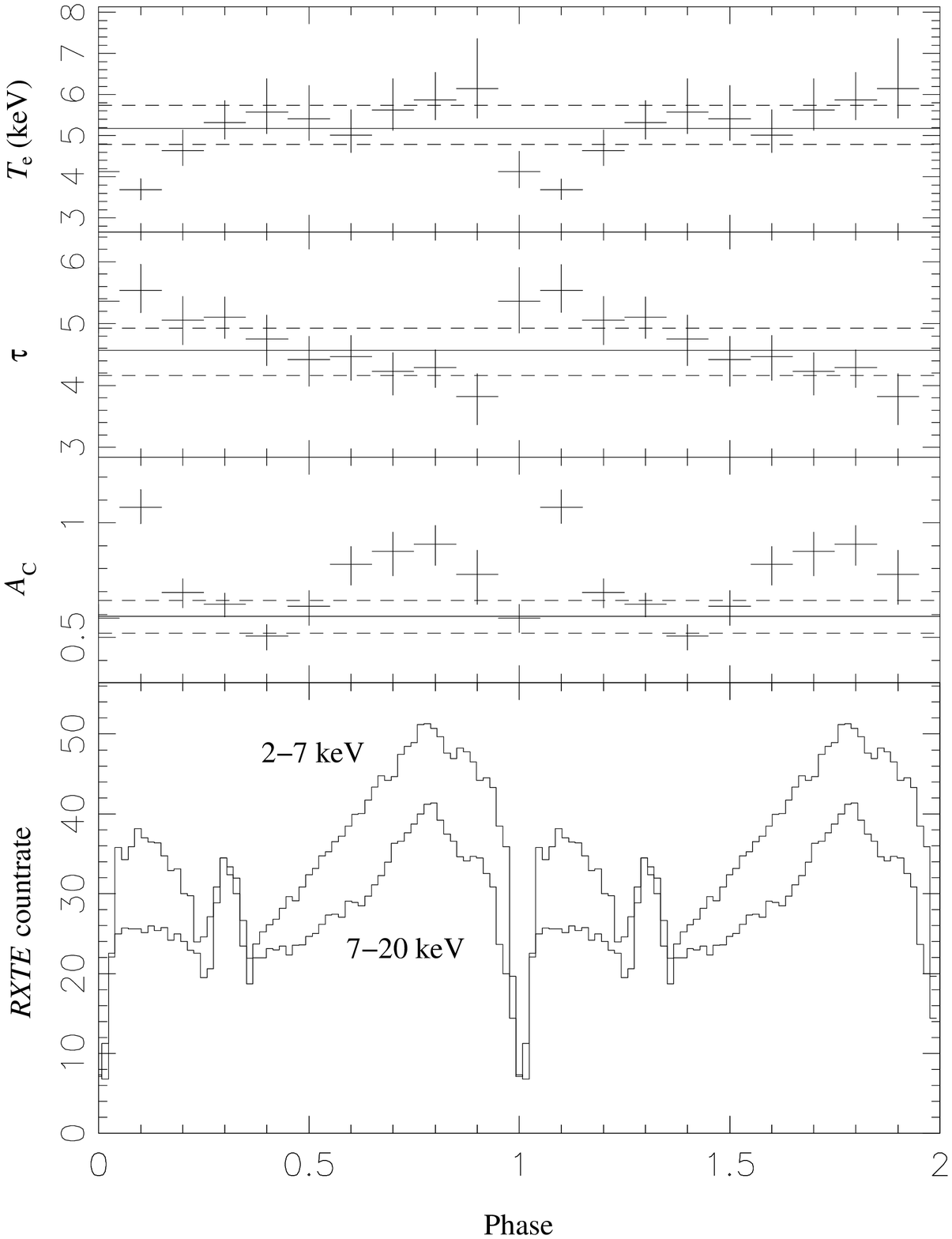}
 \caption[P40406-pps.eps]{Spectral model fitting parameters versus pulse
phase for the 1999 March 25 {\it RXTE\/} observation of RX~J0812.4$-$3114.
Phase selection was undertaken using a period $P=31.8856239$~s, with
ephemeris chosen arbitrarily so that the primary minimum falls at
$\phi=0.0$. From top to bottom, the panels shows the scattering plasma
temperature $T_{\rm e}$, optical depth for scattering $\tau$,
normalisation $A_{\rm C}$ (in units of $10^{-2}$~photons$\,{\rm
cm^{-2}\,s^{-1}\,keV^{-1}}$) and the pulse profile in 2--7 and 7--20~keV
energy bands respectively. Fitted parameter values for the phase-averaged
spectra covering the same interval are shown by the solid lines; error
bars and the dotted lines show the $1\sigma$ confidence intervals.  Two
full pulse periods are shown for clarity.  \label{P40406-pps} }
\end{figure}

Also suffering from a much lower phase-averaged countrate, spectral
fitting for the data from RX~J0812.4$-$3114 was substantially more
difficult. With all parameters free to vary, the fit values of $T_0$ and
$n_{\rm H}$ were poorly constrained, and reliable confidence limits could
not be obtained.  Following the approach of \nocite{gal00:spec}{Galloway} {et~al.} (2000), we fixed
(`froze') these two parameters at the values determined for the
phase-averaged spectra, and re-fit the pulse phase spectra.  The phase
bins centred on $\phi=0.0$ and 0.1 again show evidence for reduced $T_{\rm
e}$ and enhanced $\tau$ (Fig. \ref{P40406-pps}), although at a
significance level of only marginally greater than $1\sigma$.  While the
increase in $\tau$ coincident with the primary minimum for the latter two
observations is not significant, note that the dip is much narrower
(particularly for RX~J0812.4$-$3114) than in the 1996 February 17
observation of GX~1+4.  Thus the phase bin centered on the primary minimum
also covers the dip ingress and egress, and emission during those
intervals will tend to dominate the spectrum due to the increased
countrate.  We suggest that a spectrum taken over the phases covering only
the bottom of the dip may exhibit a much higher $\tau$, although {\it
RXTE} does not provide sufficient sensitivity in the present observations
to confirm this.

\subsection{Dip profile measurements in GX~1+4}

\begin{figure}
 \epsfxsize=9.0cm 
 \epsfbox{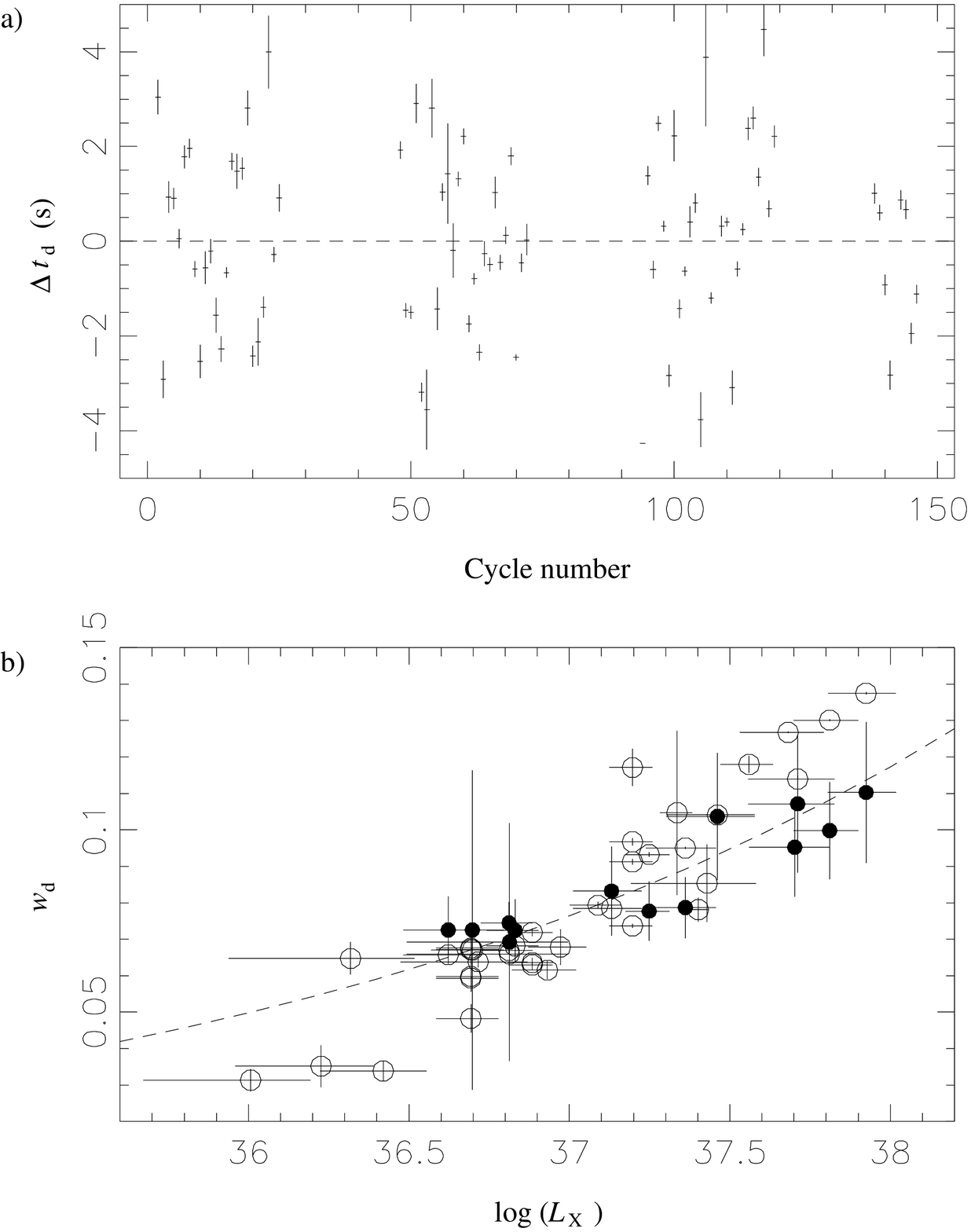}
 \caption[nature3.eps]{ Dip analysis results from GX~1+4.
a) Typical dip arrival time residuals $\Delta t_{{\rm d,}i}$ plotted
against cycle number $i$ for {\it RXTE\/} observations of GX~1+4 on 1996
February 17. The $\Delta t_{{\rm d,}i}$ from each cycle is the
difference between the expected dip arrival time (based on the mean pulse
period measured over the entire observation) and the observed time from
fitting a parabola to the lightcurve in a 16~s window centred on each
dip.  Missing cycles are a consequence of occultations of the source by
the Earth and passages of the satellite through the South Atlantic Anomaly
(SAA).
b) Dip width $w_{\rm d}$ from 35 separate {\it RXTE\/} observations of GX~1+4
between 1996 February and 1997 May and comparison with the theoretical
relationship.  The dip width calculated from the mean pulse profile is plotted
as open circles, while the averaged calculated from individual dip widths is
plotted as filled circles. The $x$-axis is the log of the 2--60~keV flux in
${\rm erg\,s}^{-1}$, assuming a source distance $d=10$~kpc. The actual 
distance
is thought to be 3--15~kpc \nocite{chak97:opt}({Chakrabarty} \& {Roche} 1997). The dashed line represents the
theoretical calculation (equation \ref{wdprop}).  Error bars represent the
1$\sigma$ uncertainties.
 \label{column} }
\end{figure}

At moderate countrates the individual dips forming the primary minimum are
usually well defined in the 1~s binned lightcurves from GX~1+4. With the
exception of the observations with very low countrates (where individual
dips could not be unambiguously located), we measured the
dip phase variations by calculating the arrival time residuals $\Delta
t_{{\rm d},i}$ for the dip from each cycle $i$. The expected arrival time
is calculated from the mean period for the observation and the ephemeris
(used to locate the first observed dip). The actual arrival time is
estimated from fitting a parabolic profile to the lightcurve in a 16~s
window surrounding the predicted arrival time; the arrival time residual
is the difference of the two.  For the 1996 February 17 observation
$|\Delta t_{{\rm d},i}|$ is typically $\sim2$~s and may be as much as 4~s
(Fig. \ref{column}a).  Visual inspection of the individual fits confirms
that the arrival time errors are genuine and not simply an artefact of
poor fits to the dip profiles (for example).  The residuals do not appear
to exhibit any periodic or quasiperiodic variation with cycle number
(time).  A search for periodic signals using the Lomb-Scargle periodgram
\nocite{nr}({Press} {et~al.} 1996) over all residuals for each of the observations on which the
analysis was performed did not result in any significant detections. The
most significant peaks were found for low frequency signals in the longest
observations, but these detections are most likely aliasing related to
observation interruptions due to earth occultations once every 90~min
satellite orbit.  The standard deviation of the dip residuals
$\sigma(\Delta t_{{\rm d},i})$ was found to vary significantly with the
mean flux.  Above $L_{\rm X}\approx4\times10^{37}\ {\rm erg\,s^{-1}}$
(2--60~keV, assuming a source distance of 10~kpc) $\sigma(\Delta t_{{\rm
d},i})$ varies between 1.4--1.9~s, whereas below that level is $\la1.3$~s
and may be as low as 0.7~s.

The dip phase width $w_{\rm d}$ was measured from the mean pulse profiles
at a countrate level midway between the minimum and the mean level in the
phase range 0.8--0.9 and 0.1--0.2. While $w_{\rm d}$ varied significantly
with $L_{\rm X}$ (Fig. \ref{column}b, open circles), the dip phase
variation described above must account for some if not all of this
dependency. Thus an equivalent measure was made using the width parameter
from the parabolic fitting divided by the mean countrate over the fitting
window, averaged over each observation. While this measurement could
not be made over the entire flux range, the resulting dip width does
indeed show less variation with $L_{\rm X}$ (Fig. \ref{column}b, filled
circles) and in particular was substantially lower at the highest
luminosity, where the dip arrival time residuals are greatest.


 \section{Discussion}

RX~J0812.4$-$3114 provides another example of an X-ray pulsar after GX~1+4
whose mean spectra are broadly consistent with Comptonisation continuum
components.  In contrast to GX~1+4, the properties of this source are
probably typical of many others, since Be/X-ray binaries make up a
significant proportion of all known X-ray pulsars.  The phase-averaged
spectral fit parameters fall within the ranges expected for neutron star
accretors \nocite{myphd:a}({Galloway} 2000b). We suggest that Comptonisation continuum
components are excellent candidates for fits to spectra from X-ray
pulsars.

The sharp dip in the pulse profiles is clearly associated with an increase
in the scattering optical depth $\tau$ both for GX~1+4 (at distinctly
different fluxes and phase-averaged spectral conditions) and also for
RX~J0812.4$-$3114. It is true that the estimated significance is only
convincing for the brightest observation of GX~1+4.  We note however
that the confidence interval limits for the pulse phase spectral fit
parameters were determined by projecting the full $p$-dimensional
confidence region (where $p$ is the number of free parameters in the model)
onto the axis corresponding to each parameter. This provides a
conservative estimate of the confidence intervals for individual parameters
\nocite{lmb76}({Lampton}, {Margon} \& {Bowyer} 1976) and so the true significance of the spectral variation may
be much greater.

Variation in $\tau$ with pulse phase is possible if the scattering region
lacks rotational symmetry, e.g. is extended in one direction
preferentially. In that case the maximum possible optical depth will be
obtained when the line of sight to the observer coincides with the longest
dimension of the scattering region.  For the accretion column, the longest
dimension is clearly along the column axis. Thus, we identify the sharp
dips as `eclipses' resulting from an approximate alignment of the
accretion column with the line of sight, occuring once each rotation
period ($\phi=0.0$ in Fig. \ref{schem}).  Absorption dips in the orbital
lightcurves of some polars, which are white dwarf stars also accreting
through a magnetically confined column \nocite{grei98}(e.g.
RX~J1802.1+1804; {Greiner}, {Remillard} \& {Motch} 1998), have previously been associated with a
similar effect. The spectral variation in the latter source during the
eclipse suggests a warm absorber, in good agreement with the results
described here.

Clearly such interactions require a fortuitous combination of the magnetic
colatitude $\beta$ (the angle between the neutron star spin axis and the
magnetic axis) and the inclination angle $i$ (between the spin axis and
the line of sight).  Since the base of the accretion column is offset from
the magnetic pole, and the column meets the surface at an angle with
respect to the local normal, $i$ must substantially greater than $\beta$
for eclipses to occur. The column orientation also depends upon the inner
disc radius $r_{\rm A,disc}$ since this defines which field lines connect
the boundary region to the neutron star.  The small number of sources
exhibiting such eclipses (3 of more than 70 known pulsars) appears to be
consistent with the relative unlikelihood of such an agreement; no
reliable estimates of $i$ and $\beta$ have been made for the three pulsars
described which might rule out this possibility.

While numerical simulations based on unmagnetized Compton scattering in a
simplified emission region support this interpretation \nocite{gal00:beam}({Galloway} \& {Wu} 2001)
it is important to note that the strong magnetic field present in the
accretion column will significantly affect the scattering cross section
\nocite{dh86}({Daugherty} \& {Harding} 1986).  The cross section $\sigma(\nu,\theta)$ will depend on both
the photon energy $\nu$ and the angle between the propagation direction
and the local magnetic field vector $\theta$.  For GX~1+4, the estimated
magnetic field strength implies an energy for the fundamental cyclotron
resonance of $h\nu_{\rm cyc}\ga200$~keV, much greater than the PCA energy
band.  Thus for a particular photon energy, $\sigma_{\nu,\theta}$ will
reach a minimum for photons propagating along the magnetic field lines
($\theta=0$). In the absence of anisotropies in the distribution of matter
this would result in enhanced emission along the magnetic axis.  This
effect will clearly compete with the decrease in emission due to the
additional optical depth experienced by photons propagating along the
accretion column.  However $\sigma_{\nu,\theta}$ reaches a minimum value
(which depends upon the magnetic field strength and the photon energy) for
photons propagating exactly along the magnetic field lines.  The increase
in optical depth is instead limited only by the geometry and the curvature
of the accretion column.  Despite magnetic anisotropy effects it should
still be possible to get a significant drop in emission due to accretion
column eclipses.

The observation of accretion column eclipses in GX~1+4 provides a unique
opportunity to study accretion dynamics under the influence of a strong
magnetic field.  The variability in the observed phase of individual dips
suggests that the column is wandering stochastically in longitude by
2--6$^{\circ}$, depending upon the source luminosity.  It is possible that the
dip timing residuals are a consequence of variations in the $P_{\rm spin}$
itself, although we consider this unlikely.  Accreting material typically
posesses a large specific angular momentum with respect to the neutron
star, and in combination with magnetic torques transmitted from the
accretion disc via the magnetic field may induce systematic period
evolution of more than 2 per cent per year.  However the variations seen in the
{\it RXTE\/} observations are typically of the same order in just one
pulse period, and in addition there appears to be no cumulative drift
in the pulse period on time--scales of tens of minutes or more as a
consequence of the timing residuals.  The accretion column is delineated
by the `bundle' of magnetic field lines which bound it, and it seems more
likely that the position of the column is varying perhaps as a consequence
of accretion along a varying set of field lines, resulting in a change in
its apparent position.

The dip phase width $w_{\rm d}$ measured from both the mean pulse profiles
and the average of the individual dip fits from each observation show a
significant correlation with X-ray luminosity $L_{\rm X}$.  This parameter
is a measure of the width of the accretion column, which ultimately
depends upon the azimuthal extent of the accretion stream at the inner
accretion disc boundary (see Fig. \ref{schem}).  The detailed physics of
this boundary region are poorly understood, and so we assume that the
entrainment of disc material to the column takes place at a rate which is
independent of the plasma density in the disc. The column will then be
wider at higher $\dot{M}$ due to the increased velocity of disc plasma as
the inner disc edge moves closer to the neutron star. The inner disc
radius is determined by a balance of the magnetic stress and the ram
pressure
\begin{eqnarray}
 r_{\rm A,disc}&\approx & 0.5\times2^{-3/7}\mu^{4/7}(GM_*)^{-
1/7}\dot{M}^{-2/7}
			\nonumber \\
            & = & 2.4\times10^8 B_{12}^{4/7} \dot{M}_{16}^{-2/7}\ {\rm cm}
 \label{alfven}
\end{eqnarray}
where $\mu$ is the magnetic dipole moment, $B_{12}$ is the corresponding
surface magnetic field strength in units of $10^{12}$~G,
$M_*=1.5\ {\rm M}_{\sun}$, $R_*=10$~km the mass and radius of the 
neutron
star and $\dot{M}=\dot{M}_{16}\times10^{16}\ {\rm g\,s^{-1}}$ the
accretion rate \nocite{gl79a,gl79b}({Ghosh} \& {Lamb} 1979a,b).  Assuming that the plasma in the
boundary region is still flowing tangentially at approximately the
Keplerian velocity $v_{\rm K}=\sqrt{GM_*/r}$ corresponding to stable
circular orbits, we have
\begin{equation}
  w_{\rm d} \propto \frac{v_{\rm K}}{r_{\rm A,disc}}
            \propto L_{\rm X}^{3/7} \label{wdprop}.
\end{equation}
The agreement between this relationship and the $w_{\rm d}$ measured from
the individual dips is encouraging (Fig. \ref{column}b), although the
error bars on several points are large indicating significant variation of
dip width for individual observations.

We note that the dip width measured for RX~J0812.4$-$3114 on 1999 Mar 25
was $w_{\rm d}\approx0.05$, comparable to that measured in GX~1+4 at
similar luminosity levels. If the conditions in the disc--magnetosphere
boundary layer are similar in both pulsars, and relative bolometric
corrections are negligible, this implies that the magnetic moments
are similar and that the field strength for RX~J0812.4$-$3114 is also
around 2--3$\times10^{13}$~G. That these two pulsars, both with atypically
strong magnetic fields, also posess an unusual alignment permitting
accretion column eclipses seems rather unlikely however.

Since the dips are associated with interactions with the accretion column,
they are expected to always be present so long as substantial amounts of
material are accreting (i.e. $\dot{M}\neq0$). The {\it RXTE\/}
observations of GX~1+4 span more than two orders of magnitude in $L_{\rm
X}$, and dips were observed in the mean pulse profiles of every
observation except one.  During a short {\it RXTE\/} observation on 1997 May
17 , the dips were weak or absent in the lightcurve even though the
phase-averaged countrate from the source was $\approx 150\ {\rm
count\,s}^{-1}$, still well above the background which is typically $100\
{\rm count\,s}^{-1}$.  The X-ray spectrum measured at this time was
unusually hard, with Compton scattering optical depth $\tau\approx19$; the
pulse-phase spectral variation was also substantially different from that
measured at other times \nocite{myphd:b}({Galloway} 2000a). Both the disappearance of the
dip and the spectral measurements indicate a distinctly different
accretion regime than is normal for this source, in which the column is no
longer present in the usual sense.

This work has revealed some important new phenomena likely to be common to
other X-ray pulsars.  Further observational and theoretical development
appears promising, in particular where the unusual properties of sources
like GX~1+4 and RX~J0812.4$-$3114 may be exploited.  The influence of the
longitudinal variation in column position on the net torque transmitted to
the neutron star may be non-negligible. This may affect the long-term
period evolution in persistent pulsars, which is generally rather poorly
understood \nocite{bil97}(e.g. {Bildsten} {et~al.} 1997).  Measurements of the dip (and hence
column) profile may provide more detailed information regarding the plasma
uptake to the accretion column at the inner disc radius. Finally, the
suggestion of disruption of the accretion column coupled with distinct
spectral states at high and low luminosities \nocite{myphd:a}({Galloway} 2000b) provides
evidence for significantly more complex flow dynamics than are predicted
by current theory.

\section*{Acknowledgments}

This research has made use of data obtained from the BATSE pulsar group
home page at http://www.batse.msfc.nasa.gov, and also the High
Energy Astrophysics Science Archive Research Center Online Service,
provided by the NASA/Goddard Space Flight Center.  The {\it RXTE\/} Guest
Observer Facility provided timely and vital help and information
throughout.




\bsp

\label{lastpage}

\end{document}